\documentclass[aps,prb,twocolumn,showpacs]{revtex4}
\usepackage[dvips]{graphicx}
\usepackage{color}

\begin{document}

\title{The effect of dispersive optical phonons on the behaviour of a
  Holstein polaron}
\author{Dominic J. J. Marchand and Mona Berciu} 

\affiliation{ Department of
  Physics and Astronomy, University of British Columbia, Vancouver,
  BC, Canada, V6T~1Z1 }

\begin{abstract}
We use the approximation-free  Bold Diagrammatic Monte Carlo technique
to study the effects of a finite dispersion of the optical
phonon mode on the properties of the Holstein polaron, especially its
effective mass. For weak electron-phonon coupling the effect is very
small, but it becomes significant for moderate and large
electron-phonon coupling. The effective mass is found to increase
(decrease) if the phonon dispersion has a negative
(positive) curvature at the centre of the Brillouin zone. 
\end{abstract}


\maketitle

{\em Introduction} --- Electron-phonon (e-ph) coupling and its effects
on the properties of quasiparticles is one of the fundamental topics
in condensed matter physics.\cite{mahan} In particular, for a very
weakly doped insulator, one can investigate the interactions of a
single carrier with the phonon distortion that builds up in its
presence, to understand the properties of a single polaron -- the
dressed quasiparticle consisting of the carrier and its phonon cloud.

Most such polaron studies focus on e-ph coupling to optical phonons. The
reason is that (assuming that long-range interactions are 
screened) the strength of the short-range coupling depends on the
relative displacements between the atom hosting the carrier and its
neighbours, because these displacements modulate the on-site energy
and hopping integrals of the carrier. Acoustic phonons are gapless and
thus easy to excite, however, since they describe ``in-phase'' motion
of neighbouring atoms,\cite{mahan} the corresponding relative
displacements are vanishingly small leading to very weak e-ph
coupling. In contrast, optical phonons describe ``anti-phase'' atomic
motion\cite{mahan} which leads to large relative displacements and
therefore a much stronger e-ph coupling.

Aside from ignoring coupling to acoustic phonons,\cite{marsiglio} another
widespread approximation is to assume that the optical phonon is
dispersionless (Einstein model).\cite{raedt} In part, this is due to
the belief that this should be a good approximation if the phonon
bandwidth is small compared to the average phonon energy.\cite{comm}
In practice, this is also due to the scarcity of numerical techniques
suitable for the study of polaron models with dispersive
phonons. These reasons explain why very little is known about the
effect of dispersive optical phonons on polaron properties. For
example, all else being equal, will dispersive phonons increase or
decrease the polaron's mass? The fact that the phonons have a finite
speed might suggest the former, since the cloud is now mobile.
However, mobile phonons will spread in all directions away from the
carrier; this may lead to a more extended and thus harder to move 
cloud (for a fixed number of phonons). 
Apart from clarifying qualitative trends, it is important to
understand quantitatively the accuracy of this approximation, as it
may have important consequences in the modeling of materials and of
quantum simulators.\cite{qsims}

In the wake of the discovery of several instances where what were long
believed to be standard polaronic properties were found to break down
upon straightforward extensions of polaronic
models,\cite{PRL105.266605, EPL102.47003} we believe it is high time to
reexamine  this approximation of an Einstein mode.\cite{raedt} Here,
we investigate the effect of dispersive 
phonons in the most well understood polaron model with
short-range e-ph coupling,\cite{revs} the Holstein model.\cite{holstein}

{\em Model and methods} -- We consider the generalized Holstein model for a
one-dimensional chain with $N\rightarrow \infty$ sites and lattice
constant $a=1$:
\begin{eqnarray}\label{eqn_Holstein_H}
\hat{\mathcal{H}}&=&\sum_{k}^{}\epsilon(k) c_{k}^\dagger
c_{k}+\sum_{q}^{}\Omega(q) b_{q}^\dagger b_{q} \nonumber
\\ &\phantom{=}& + \frac{g}{\sqrt{N}} \sum_{k,q}^{}c_{k+q}^\dagger
c_{k} \big( b_{-q}^\dagger +b_{q} \big),
\end{eqnarray} 
where $c_k$ annihilates a carrier with momentum $k$ (the spin degree
of freedom is trivial so it is not listed) while $b_q$ annihilates a
phonon with momentum $q$. Sums are over the Brillouin zone $-\pi < k
\le \pi$. The first term is the carrier's kinetic energy with the
usual tight-binding dispersion
$\epsilon(k) = -2t\cos(k)$;  $t$ is the hopping constant. The
second term describes the  phonons. For simplicity, we
forgo an exact description of the optical phonon dispersion for a
detailed model of a lattice with a non-trivial basis, and instead use:
\begin{equation}\label{eq_wq}
\Omega(q) = \Omega_0 - \Delta \Omega/2 \big[ 1-\cos(q) \big].
\end{equation}
$\Omega_0=\Omega(0)$ is the long-wavelength limit of the phonon
energy. Since $\Omega(\pi)=\Omega_0-\Delta \Omega $, the phonon
bandwidth is $\Delta\Omega$. We will focus on
downward (normal) dispersion with $0 < \Delta \Omega < \Omega_0$, but
will also comment on the case with $\Delta \Omega <0$. The third term
is the Holstein coupling.\cite{holstein}

Taken at face value, this Hamiltonian  is  unphysical:
a chain with one atom per unit cell does not support optical
phonons. One should think of the orbitals kept
in the model as being the subset of valence orbitals associated with a
more complex unit cell, with the assumption that the other orbitals
are located at energies far enough removed from these that their
occupation numbers cannot change. From a pragmatic point of view,
our motivation is to study a polaron model that is very well
understood for  dispersionless phonons, to see the effects
of changing only this assumption.

The properties of the Holstein polaron for Einstein phonons ($\Delta
\Omega=0$) are controlled by two dimensionless parameters: (i) 
the effective coupling $\lambda=g^2/(zt\Omega_0)$ equal to the
ratio of the deformation energy $-g^2/\Omega_0$ (the polaron
energy in the impurity limit $t=0$) to the free electron energy
$-zt$,  $z$ being the coordination number ($z=2$ in 1D).  For
$\Omega(q)$ of Eq. (\ref{eq_wq}) the deformation energy is
$-g^2/\sqrt{\Omega_0(\Omega_0-\Delta \Omega)}$ (see below), thus:
\begin{equation}\label{eqn_lambda}
\lambda = \frac{g^2}{2t\sqrt{\Omega_0(\Omega_0 - \Delta \Omega)}},
\end{equation}
and (ii) the adiabaticity ratio $\Omega_0/t$.
We round up these parameters with (iii)  the dimensionless phonon bandwidth
$\delta = \Delta \Omega/\Omega_0< 1$.  Hereafter we set  $t=1$.

We investigate the effect of  $\Delta \Omega \ne 0$ on  the polaron
energy $E(k)$, {\em i.e.} the energy of the lowest eigenstate of
momentum $k$,  $\hat{\mathcal{H}}| \tilde{k}
\rangle =E(k) | \tilde{k}
\rangle$, 
and its quasiparticle weight $Z(k)=|\langle k | \tilde{k} \rangle|^2$
given by the overlap between the 
polaron eigenstate $| \tilde{k} 
\rangle$ and the free electron state $| k \rangle= c_k^\dagger|0\rangle$.
We also study the effective
 polaron mass $m^*/m_0 = 2t / \frac{\partial^2E(k) }{\partial k^2}|_{k=0},$
%
where the bare electron mass is $m_0=1/2t$.

The results presented here are obtained with a variant of the
Diagrammatic Monte Carlo (DMC) technique \cite{JETP87.310,
  PRB62.6317,PRL81.2514} known as the Bold Diagrammatic Monte Carlo
(BDMC) technique.\cite{PRB77.125101, PRL105.266605} Like DMC, BDMC
consists in a Monte Carlo sampling of the Feynman diagrammatic
expansion of the continuous imaginary-time self-energy of the
polaron. The difference is that this sampling is done with the
electron bare propagator self-consistently replaced by a dressed
propagator as the calculation progresses (thus speeding up
convergence), while enforcing necessary restrictions on the topology
of the diagrams to avoid double-counting. As is the case for DMC, BDMC
is exact within the limits of its statistical error bars and does not
make any assumptions or enforce any non-physical restrictions. DMC and
its variants can treat dispersive (including acoustic)
phonons efficiently,   provided we avoid the extreme
adiabatic regime $\Omega_0/t\ll1$.

\begin{figure}[t]
        \includegraphics[width=0.98\columnwidth]{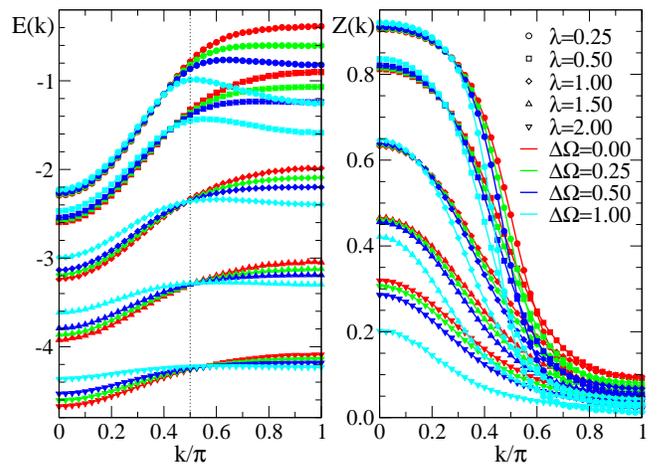}
\caption{(Color online) Polaron energy $E(k)$ and
  quasiparticle weight $Z(k)$ for $t=1, \Omega_0=2$ and various phonon
  bandwidths (red, green, blue and cyan are for
  $\Delta \Omega=0.0, 0.25, 0.50$ and $1.00$, respectively) and effective
  couplings $\lambda$ (with 
  circles, squares, diamond, up-triangles and down-triangles for
  $\lambda=0.25, 0.50, 1.00, 1.50$ and $2.00$, respectively).} \label{fig1}
\end{figure}

{\em Results and Discussion} --- Figure \ref{fig1} shows BDMC results
for  $E(k)$ and  $Z(k)$  for increasing $\Delta
\Omega/\Omega_0$ and different 
values of $\lambda$. In all cases $t=1,
\Omega_0=2$. The error bars are smaller than the size of the symbols.

Consider first $\Delta \Omega=0$, {\em i.e.} the usual Holstein
model. As expected,\cite{revs} both $E(k)$ and $Z(k)$ are monotonic
functions of the momentum $k$.  With increasing $\lambda$, the polaron bandwidth
$E(\pi)-E(0)$ becomes narrower, signalling an increasingly heavier
polaron;  the quasiparticle weight decreases
considerably, even at $k=0$. Note that for
this value of $\Omega_0$, even 
$\lambda=2$ is still in the intermediary regime, with a ground-state
quasiparticle weight $Z(0)\approx 0.32$. 

We can now gauge the effects of increasing  $\Delta
\Omega$. Starting first with $E(k\sim 0)$, we see that this
results in an increase in the polaron energy (for the same value of
$\lambda$) and a significant additional flattening of the
band, implying an even heavier effective mass (see
below). Quantitatively, both these effects increase with increasing
$\lambda$. 

Near the edge of the Brillouin zone, we see a
downturn of the dispersion relation which is more pronounced for
smaller $\lambda$. This downturn is due to the fact that the
polaron band must lie below the polaron+one-phonon
continuum, which comprises excited states where the polaron scatters
on one or more phonons that do not belong to its cloud. The lower edge
of this continuum is at $\min_q[E(k-q) + \Omega(q)]$. If $\Delta
\Omega =0$, this continuum starts at
$E(0)+\Omega_0$ for all $k$. If $\Delta \Omega \ne 0$, its 
boundary varies with $k$. In particular, for $k=\pi$
and normal (negative) phonon dispersion, the minimum is reached at
$E(0)+\Omega(\pi)$ where both terms are minimized. This agrees with
the results of Fig. \ref{fig1} and explains the apparent ``folding''
of the polaron dispersion. One immediate conclusion is that for
dispersive phonons, one cannot use the location of the continuum to
fully determine the phonon energy. 

The effect of increasing $\Delta \Omega$  on the $Z(k)$ is different
in the weak and strong coupling regimes, for 
$k=0$. Here we see an increase of $Z(k\approx 0)$ with $\Delta \Omega$
at small coupling, while for stronger coupling $\lambda$ it decreases
with increasing $\Delta \Omega$. As we show below, the effective 
mass increases with increasing $\Delta \Omega$ for all
$\lambda$. This warns against inappropriate use of the  familiar
relationship,  $m^*/m_0=1/Z(0)$, which only
holds if the self-energy is momentum
independent.\cite{mahan} For the usual Holstein model with $\Delta
\Omega=0$, the momentum dependence of the self-energy is so weak
that this equality is satisfied with reasonable accuracy. For
dispersive phonons, however, the self-energy acquires significant
momentum dependence (the downturn in $E(k\approx \pi)$ is another
indication that this must be the case). As a result,  $m^*/m_0=
1/Z(0)$ is no longer valid. 

\begin{figure}[t]
        \includegraphics[width=0.98\columnwidth]{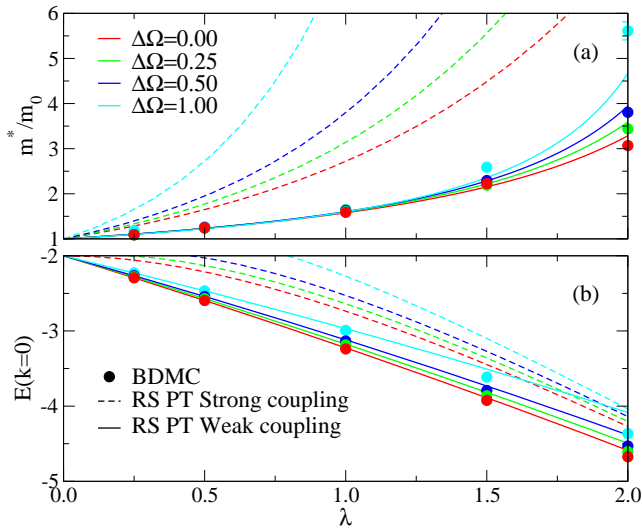}
\caption{(Color online) (a) Effective mass $m^*/m_0$, and (b) 
  polaron ground-state energy $E(0)$ vs. the effective
  coupling $\lambda$ for various phonon bandwidths $\Delta
  \Omega$. Symbols show BDMC results, with error bars smaller than the
symbol size except where explicitly shown. Full lines are results from
Rayleigh-Schr\"odinger second-order (i.e., fourth order in the
electron-phonon coupling) perturbation theory for weak coupling, while
dashed lines are results from Rayleigh-Schr\"odinger first order
perturbation theory for strong coupling.
} \label{fig2}
\end{figure}

Figure \ref{fig2}(a) plots $m^*$  vs.
$\lambda$ for several  values of  $\Delta
\Omega$. Symbols show $m^*$ extracted from  $E(k)$ by fitting it to
$\sum_{n\le 6} a_n\cos(nk)$. We find 
that $m^*$ increases with increasing $\Delta \Omega$ 
for all $\lambda$, although for small $\lambda$ the change is very
small. To confirm this, we use perturbation theory.

As is usually the case for polaron problems, we find that the
Rayleigh-Schr\"odinger approach  yields better agreement
than  Wigner-Brillouin perturbation theory;  we restrict
ourselves to presenting results only from the former. 
For $\lambda \rightarrow 0$, we start with the free electron in a
phonon vacuum, $|k\rangle$, and take the e-ph term as the weak
perturbation. 
The first correction to the energy is:
 \begin{equation}\label{eq_rs1st_correction}
E^{(2)}(k)=\frac{1}{N}\sum_{q} \frac{g^2}{\epsilon(k) - \epsilon(k-q)
  - \Omega(q)}, 
\end{equation}
which for the dispersion of Eq. (\ref{eq_wq}) gives
\begin{equation}\label{eqn_E2_weak}
E^{(2)}(k) = \frac{-2t\lambda \Omega_0
  \sqrt{1-\delta}}{\sqrt{4t\Omega_0\cos{k} + \Omega_0^2
    (1-\delta)-4t^2\sin^2 k}}. 
\end{equation}
From this, we find:
\begin{equation}\label{eqn_m_weak}
\frac{m^*}{m_0} = \Bigg[ 1 - \frac{\lambda t \Omega_0  (2\Omega_0 +
    4t)\sqrt{1-\delta}}{\big[ 4t \Omega_0 + (1-\delta)
      \Omega_0^2\big]^{3/2}} \Bigg]^{-1}. 
\end{equation}

The next correction $E^{(4)}(k)$ is calculated similarly (we do not
write its long expression here). Comparison between  BDMC (symbols)
and perturbational results  $E_{\lambda \rightarrow 0}(k) = \epsilon(k) + E^{(2)}(k) +
E^{(4)}(k)$ (full lines) is shown in Fig. \ref{fig2}(b). The agreement
is good up to $\lambda \sim 1$ and then becomes progressively worse. We
note that  although the perturbational expression works very well at
$k\approx 0$ and small $\lambda$, it should not be trusted at larger
$k$ because the denominator of Eq. (\ref{eq_rs1st_correction})
vanishes at a finite  $k$, resulting in unphysical
behaviour. For  $k\rightarrow 0$, however, we 
can use it to calculate the perturbational prediction for $m^*$, shown
by full lines in panel (a). It confirms that  $m^*$ increases with
$\Delta \Omega$ although for small 
$\lambda$ the effect is tiny.
It is also  straightforward to find
$Z(k)=1- \alpha_k + \dots $ and the average number of phonons 
$N_{ph}(k) = \alpha_k + \dots$, where $
\alpha_k =g^2/N\sum_{q} \big[\epsilon(k) -
    \epsilon(k-q) - \Omega(q)\big]^{-2}$.

For $\lambda \gg 1$, we take the electron
kinetic energy as the small perturbation.\cite{PhysicaC244.21} For
$t=0$ the  ground-state is found 
using the unitary transformation $e^S$ where
\begin{equation}
S = -\sum_q \frac{g}{\Omega_q} \big[ b^\dagger_{-q} - b_q
  \big]\frac{1}{\sqrt{N}}\sum_{k}^{}c^\dagger_{k+q} c_k. 
\end{equation}
For $\Delta \Omega=0$ , this is the Lang-Firsov
transformation.\cite{JETP16.1301}   If the electron is 
located at site $i$, the transformed
phonon annihilation operator is found to be:
\begin{equation}
  B_{q,i}=\left. e^S b_q e^{-S}\right|_{\hat{n}_i=1}=b_q +
  \frac{g}{\Omega(q) } \frac{e^{-i qR_i}}{\sqrt{N}}. 
\end{equation}
Within this Hilbert subspace the Hamiltonian becomes:
\begin{equation}\label{eq_H_imp_limit}
\left.\hat{\mathcal{H}}_{t=0}\right|_{\hat{n}_i=1}=\sum_{q}^{}\Omega(q)
B^\dagger_{q,i}B_{q,i} -
\frac{g^2}{\sqrt{\Omega_0 (\Omega_0 - \Delta \Omega)}}.
\end{equation}
The second term gives the ground state energy in this limit. Its
wavefunction is  
$|\tilde{i}\rangle =c_i^\dagger \exp[-{g^2\over
    2N}\sum_{q}^{}[\Omega(q)]^{-2 }-{g\over \sqrt{N}}\sum_{q}^{}
 e^{-i qR_i}[\Omega(q)]^{-1}b_q^\dagger  ]|0\rangle$. Hopping
lifts the degeneracy between these states and, to first order, leads
to eigenstates $|\tilde{k}\rangle = {1\over \sqrt{N}} \sum_{i}^{}
e^{ik R_i} |\tilde{i}\rangle$ of energy
$ E_{\lambda \rightarrow \infty}(k) = -2t\lambda -2t^* \cos(k)+ \dots$,
where the effective hopping  $t^*$ is exponentially
suppressed so that:
\begin{equation}
\label{m}
\frac{m^*}{m_0} = \frac{t}{t^*} = e^{\frac{2t\lambda}{\Omega_0 (1 - \delta)}}+\dots  ,
\end{equation}
while $Z(k)\approx \exp[-N_{ph}]$ where the average number of phonons
is   $N_{ph} = [2 + \delta/(1-\delta)]t \lambda/\Omega_0+\dots.$

The strong-coupling perturbational results are shown as
dashed-lines in Fig. \ref{fig2}. While trends are
correct, even for $\lambda=2$ the agreement is  poor. 
This is not surprising because, as mentioned, $\lambda=2$ is only
an intermediate coupling for the adiabaticity ratio used
here. Eq. (\ref{m}) confirms that $m^*$ increases with $\Delta \Omega$
for a fixed $\lambda$. We also see that $m^*/m_0 \ne 1/Z(0)$ even when
$\lambda \rightarrow 
\infty$.  

These results suggest a reason for the larger polaron effective mass
for (normally) dispersive optical phonons with $\delta=\Delta
\Omega/\Omega_0 >0$. The speculation that phonon mobility may lead to
a more extended cloud and thus larger $m^*$ is not borne out: phonons
are mobile for any $\delta \ne 0$ yet BDMC results (not shown) and the
perturbational formulas show that $m^*$ decreases if $\delta <0$. The
difference is that if $\delta >0$, the negative phonon dispersion
leads to a phonon speed with a sign opposite to that of the
carrier. For a polaron with small momentum $k >0$, since the carrier
must be close to its ground-state, $k_c \approx 0$, the contribution
from the momenta of all the phonons in the cloud, $\sum_{i}^{} q_i = k
- k_c $, must be small and positive. However, phonons with small
positive momentum move in a direction opposite to that of the polaron,
slowing it down (aside from being energetically costly). Less expensive
phonons with positive speed have momenta just above $-\pi$, but
balancing many such momenta to obtain a small positive total is
challenging. If $\delta <0$, however, phonons with small momentum move
in the same direction as the carrier and are the least costly, so
$m^*$ decreases with $| \delta|$.

This explanation is also supported by the fact that $m^*$ starts to
change considerably only once the polaron bandwidth becomes 
smaller than the phonon bandwidth. If the phonons are much
slower than the polaron then the cloud will primarily move through
phonon emission and absorption by the carrier, like for $\delta =0$,
and $m^*$ should not be much affected. Indeed, this is what we observe
for small $\lambda$. This also suggests that, as far as
$m^*$ is concerned, $\Delta \Omega/t^*$ may be a more suitable
dimensionless parameter to characterize the effects of phonon
dispersion. However, this may not be true for all quantities so
we continue to use $\delta$ as the third parameter.

For the usual Holstein model it is known that $\lambda$ is the primary
parameter that determines polaron properties, for example the
crossover to a small polaron occurs at $\lambda \sim 1$.\cite{revs}
The adiabaticity ratio influences the ``sharpness'' of the crossover
and its precise location but does not lead to qualitative changes, at
least not while $\Omega_0/t$ is away from the strongly adiabatic
regime (which is not suitable for study with BDMC, in any event).

Our results suggest that the same conclusions apply for a finite
phonon bandwidth $\delta$ if it is small enough to keep the phonons
gapped (as is reasonable for optical phonons\cite{comm2}). For gapped
phonons, the polaron crossover is known to remain smooth even if
$\delta \ne 0$,\cite{lowen} so $\delta$ may only affect its
sharpness. We can estimate this by considering, for example, its
influence on the evolution of $\ln m^*/m_0$ with $\lambda$. The slope
of this quantity for both $\lambda\ll 1$ and $\lambda\gg 1$ can be
obtained from perturbation theory. The larger the mismatch between the
two, the sharper must be the crossover.  If $\Omega_0/t\gg 1$, both
Eqs. (\ref{eqn_m_weak}) and (\ref{m}) give a slope of
$2t/[\Omega_0(1-\delta)]$, consistent with a very smooth crossover for
both Einstein ($\delta=0$) and gapped ($\delta<1$) dispersive phonons. 
For $\Omega_0/t\ll 1$, the slope remains the same for $\lambda \gg 1$, while 
for $\lambda \ll1$ we now find $\sqrt{t(1-\delta)/(4\Omega_0)}$. The mismatch 
is severe, explaining the expected sharp crossover in this limit for $\delta=0$. 
If $0< \delta <1$, the mismatch is further accentuated, leading to an increasingly
sharper crossover with increasing $\delta$. Fig. \ref{fig2} shows the beginning of 
this crossover, especially when combined with the $\lambda \gg 1$ results 
(dashed lines).

Similar considerations allow us to speculate on the effects of
$\delta$ for a dispersion similar to Eq. (\ref{eq_wq}) 
in $D>1$.  Again, no significant changes with $\delta$ are
expected if $\lambda$ is small. For $\lambda \gg 1$, perturbation
theory confirms an increase of $m^*$ with $\delta$ and a sharpening of the 
crossover for $\Omega_0/t\ll 1$, even if these effects are slightly weaker
than in 1D. For detailed quantitative investigations in 2D and
3D it is probably best to use a realistic phonon dispersion.

To conclude, we studied the effect of dispersive optical phonons
on the properties of a Holstein polaron. For weak
e-ph coupling, $m^*$ is little affected so
the Einstein mode approximation is valid, although it may
miss important higher energy physics such as the 
downturn in $E(k)$ at the Brillouin zone edge. For larger e-ph
coupling, however,  $m^*$ may increase (decrease) significantly if
$\delta >0$ $(\delta <0)$. This suggests that quantitative modelling of
materials with strong e-ph coupling needs to  explicitly take the
dispersion of the optical phonons into account.

{\em Acknowledgements:} We thank N. Prokof'ev for help with BDMC, and
NSERC and CIFAR for funding.


\begin{thebibliography}{99}
\bibitem{comm2}  Of course, one may
wonder what happens  when the gap is closed, but for
Eq. (\ref{eq_wq}) this 
results in gapless phonons with $\Omega(\pi-q) \sim q^2$ for $q\ll1$.
This is very different from acoustic phonons with $\Omega(q) \sim q$
for $q\ll1$ so it is not clear what would be the physical
relevance of such results.
\end{thebibliography}
\end{document}